\documentclass[prl,showpacs,aps,twocolumn]{revtex4}
\usepackage{amsmath}
\usepackage{epsfig}
\begin{document}

\bigskip
\title{Scanning the critical fluctuations \\ -- application to the phenomenology of the two-dimensional XY-model --}

\author
{Ricardo Paredes V.$^{\dagger, \ddagger}$
and
 Robert Botet$^{\star}$
}
 
\affiliation{
$^{\dagger}$ Instituto Venezolano de Investigaciones Cient\`{\i}fica, Centro de F\`{\i}sica, 
Laboratorio de Fisica Estadistica Apdo21827 
1020A  Caracas, Venezuela \\
$^{\ddagger}$ NanoStructured Materials, Delft University of Technology,
Julianalaan 136, 2628 BL Delft, Netherlands\\
$^{\star}$
Laboratoire de Physique des Solides B\^{a}t.510, CNRS UMR8502 - Universit\'{e} Paris-Sud,
Centre d'Orsay, F-91405 Orsay, France  
}
 
\date{\today}
 
\begin{abstract}
\parbox{14cm}{\rm 
We show how applying field conjugated to the order parameter, 
may act as a very precise probe to explore
the probability distribution function of the order parameter.
Using this `magnetic-field scanning' on large-scale numerical simulations
of the critical 2D XY-model, we are able 
to discard the conjectured double-exponential form of the large-magnetization 
asymptote.  
}
 
\end{abstract}
\bigskip 
\maketitle

\subparagraph*{Introduction. --} 
Derivation of the complete equation of state of a many-body system is generally a formidable task.
When the system may appear under various phases at the thermodynamic equilibrium, this problem
requires knowledge of the
exact probability distribution function (PDF) of its order parameter.
Despite a number of attempts, just a few instances 
are available \cite{Kolmogorov}. Even the exact PDF for the 2D Ising model
is still unknown. 

Within this context, the critical point is very particular, since the universality concept
tells us that only a limited information is needed to obtain the complete leading 
critical behavior. For instance, general arguments
give precisely the tail of the critical PDF, $P(m)$, 
for the large values of the order parameter, $m$, namely \cite{Bouchaud}:
\begin{equation}
P(m) \sim e^{-c m^{\delta+1}} ~, \label{tail}
\end{equation}
with $c$ a positive constant and $\delta$ the magnetic field critical exponent,
or the distribution of the zeros of the Ising partition function in the complex magnetic 
field \cite{Yang-Lee} (such a partition function is Fourier transform of the PDF). 

In the present
work, we explain how the real magnetic field can be generally used as a very accurate probe to 
scan quantitatively the zero-field PDF tail,
exemplifying the method with the critical 2D XY-model. By the way, we will see that the popular 
double-exponential approximation
of the PDF for this system cannot be correct, and we provide alternative approximation which is 
consistent with the critical behavior. Consequently, our results discard possible 
fundamental connexion between this magnetic model and the field of extremal-values statistics.

\subparagraph*{Former approximation of the magnetization PDF for the critical 2D XY-model. --}
In a series of recent papers \cite{Bram1, Bram2, Bram3, holds}, it was argued that the 
PDF $P(m)$ of the magnetization $m$ of the 2D XY-model at
the Berezinskii-Kosterlitz-Thouless (BKT) critical temperature, 
could be approximated  by the generalized  Gumbel form:
\begin{equation}
\label{bramwell}
P(m) \propto 
\exp \left( b_{\sigma} z_{\sigma} - \lambda_{\sigma} e^{a_{\sigma} z_{\sigma}} \right) ~,
\end{equation}
where the reduced magnetization: $z_{\sigma}=(m-\langle m \rangle)/\sigma$ is used. 
From low-temperature spin-wave theory and direct numerical simulations, one obtains \cite{Bram2}:
\begin{equation}
a_{\sigma}  \approx  1.105  ~~;~~
b_{\sigma}  \approx  1.74  ~~;~~
\lambda_{\sigma}  \approx  0.69   ~~. \label{asigma}
\end{equation}
It was regularly noticed \cite{Bram2} that the form \eqref{bramwell} {\em cannot} be the exact
solution of the corresponding statistical problem, even if a number of analytical arguments
as well as numerical simulations show convincingly that this trial function is indeed close to the exact 
solution. Moreover,
Eq.\eqref{bramwell} is appealing, as it suggests connexion between the critical 2D  XY-model 
and the statistics of extreme variables \cite{extremes}. Therefore, the question of a possible bridge
between these two active fields of statistical physics should be examined precisely.
On the other hand, Eq.\eqref{bramwell} is
inconsistent with the general behavior \eqref{tail}, since 
$\delta = 15$ for the 2D XY-model. The question to know whether relation \eqref{tail} is true or wrong for
this system, is then fundamentally important. 
We will examine hereafter these two questions.

\subparagraph*{Two alternative hypothesis. --}
We consider the 2D XY-model \cite{Bere} on a square
lattice of size $L \times L$ with periodic boundary conditions. 
The $N = L^2$ classical spins are confined in the $x$-$y$ lattice plane, and they interact
according to the Hamiltionian: 
$H = -J \sum_{<i,j>} {\bf S}_i \cdot {\bf S}_j$, 
where $J>0$ is the ferromagnetic coupling constant and
the sum runs over all nearest-neighbor pairs of spins.  
Eventual critical features are characterized by the singular behavior of the scalar 
magnetization per site: $m \equiv \frac{1}{N}\sqrt{\left( \sum_i {\bf S}_i \right)^2}$,
which is a non-negative real number.
We define also the instantaneous magnetization direction as the angle $\psi$ such that:
$\sum_i S_i^{x} = mN \cos \psi$ and $\sum_i S_i^{y} = mN \sin \psi$.

\begin{figure}[ht]
\begin{center}
\includegraphics[width=0.9\columnwidth]{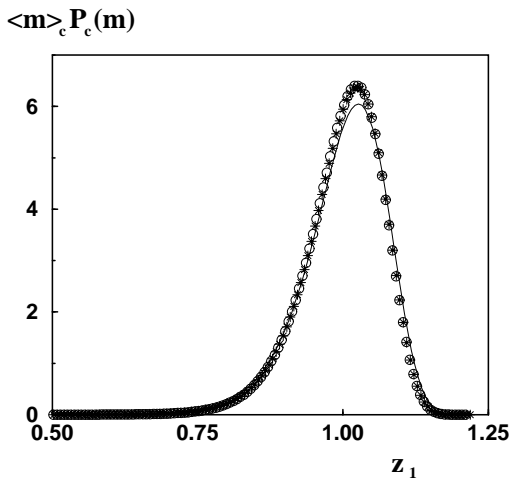}
\end{center}
\caption{\label{1stlaw} PDF of the magnetization for the 2D XY-model
at the critical temperature $T_{BKT}$, plotted in the first-scaling form \eqref{1st}.
The scaling law is confirmed for $L=64$ (stars) 
and $L=128$ (circles), while the $L=16$  (continuous line) shows  finite-size 
deviation. Wolff's single-cluster algorithm was used \cite{Wolff}. 
Each data set corresponds to average over $25,000,000$ independent realizations.}
\end{figure}
There is a continuous line of critical points 
for any temperature below the critical BKT temperature
$T_{BKT}$ \cite{KT}. In this region, $0 \le T \le T_{BKT}$, 
the system is critical, and asymptotic ({\it i.e.}  
$L\rightarrow \infty$) self-similarity
results in the so-called first-scaling law \cite{nous}:
\begin{equation}
\label{1st}
\langle m \rangle P(m) = \Phi_T(z_1) \hskip 0.75cm ~~\mbox{, with}~~~ z_1 \equiv \frac{m}{\langle m \rangle} ~ ,
\end{equation}
and $\Phi_T$ is a scaling function which depends only on the actual temperature $T$.
Under this form,
the hyperscaling relation, $\langle m \rangle/\sigma = cst$, is automatically realized. 
Eq.\eqref{1st} is a sequel of the standard finite-size scaling theory \cite{FFS},
but it is highly advantageous 
that \eqref{1st} does not require knowledge of any critical exponent.
FIG.(\ref{1stlaw}) gives numerical exemplification of the first-scaling law at $T_{BKT}$,
and illustrates the overall shape of the distribution $\Phi_{c}(z_1)$
(hereafter, the index `$c$' refers to the BKT critical point, $T=T_{BKT}$). \\

\begin{figure}
\begin{center}
\includegraphics[width=0.9\columnwidth]{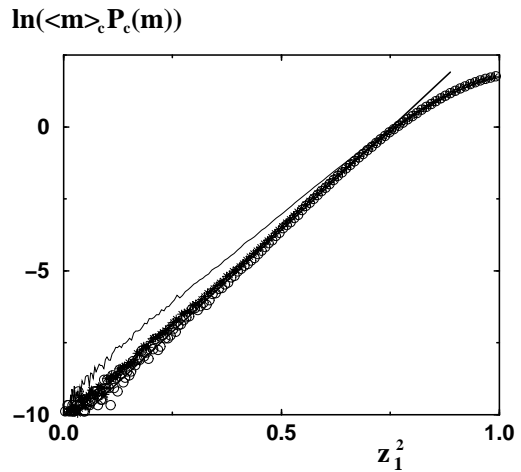}
\end{center}
\caption{\label{TKT} Part $m \leq  \langle m \rangle_c  $ of the logarithm of the 
scaled PDF \eqref{1st} {\it vs} $z_1^2$, for the 
2D XY-model at $T=T_{BKT}$. The solid straight line is the best fit:
$\ln (\langle m \rangle_c P_c(m) ) = b_1 z_1^2+$cst 
for the $L=128$, $z_1<0.8$, data. Numerically, $b_1 = 12.7$. Same symbols as in Fig.\ref{1stlaw}.}
\end{figure}

We separate the free energy ${\cal F}$ of the 2D XY-system at equilibrium
(temperature $T=1/\beta$) into the sum of a regular part describing the small values of
the magnetization, a singular part \cite{Widom} vanishing 
as the essential singularity \cite{Amit, Gulacsi} 
when $T \rightarrow T_{BKT}$, and
a regular part for the large values of the magnetization, namely:
\begin{equation}
\label{free}
\beta {\cal F}(m) = \varphi_0(m/\langle m \rangle)+\varphi_S(m/\langle m \rangle)+
\varphi_{\infty}(m/\langle m \rangle) ~.
\end{equation}
Clearly, discussion on the
system behavior can be carried out either through the free energy 
\eqref{free} or the first-scaling law \eqref{1st}, since: $\ln P(m) = -\beta {\cal F}(m) +  $constant term.
\paragraph*{The regular small-$m$ tail. --}
As the  singular behavior should vanish at the BKT transition,
we study first the  regular small-$m$ behavior of $P(m)$
at $T=T_{BKT}$. Numerical results for $P_c(m)$ are shown on FIG.\ref{TKT} in the
form \eqref{1st}. They suggest the leading form: 
\begin{equation}
\label{correction}
\ln P_c(m) \approx b_1 \left( m / \langle m \rangle_c \right)^2 ~ .
\end{equation}
Close to the most probable value of the magnetization, Eq.(\ref{bramwell}) and Eq.(\ref{correction}) 
are indeed consistent each other as the latter writes : 
$\ln P_{c}(m) = 
\mbox{cst} + 2 b_1 z_{\sigma} / (\langle m \rangle_c / \sigma_c) + {\cal O}(z_{\sigma}^2)$, 
in which we recognize the term linear in $z_{\sigma}$.

\begin{figure}[ht]
\begin{center}
\includegraphics[width=0.9\columnwidth]{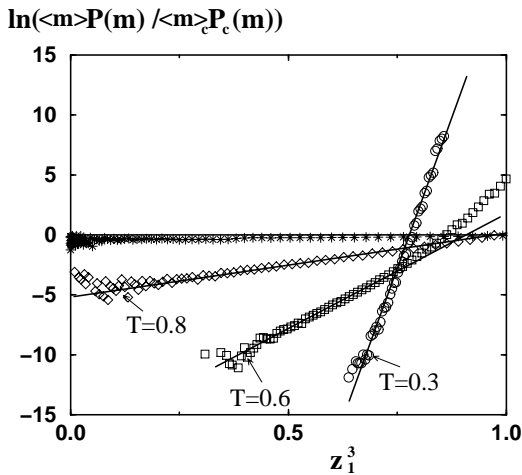}
\end{center}
\caption{\label{corr} Part $m \leq  \langle m \rangle  $ of the logarithm of the 
scaled PDF, corrected by 
the regular part of the free energy (see (\ref{corrige})), for $L=16$ and 
four different temperatures: $T=0.3$ (circles), 
$T=0.6$ (squares), $T=0.8$ (diamonds), and $T=0.885$ (stars) which is close to the 
critical temperature ($T_{BKT} \approx 0.893$ \cite{Berche}). 
The plot is versus $z^3_{1} \equiv (m/\langle m \rangle)^3$. 
The straight lines are the best  fits Eq.\eqref{singul}.}
\end{figure}

\paragraph*{The singular small-$m$ tail. --}
We consider now the singular part of the free energy through the combination:
\begin{equation}
\label{corrige}
\ln \left( \langle m \rangle P(m) \right) - \ln \left( \langle m \rangle_c P_c(m) \right)
\end{equation}
{\it vs} the reduced magnetization $z_{1} \equiv m/\langle m \rangle$.
The data plotted in 
FIG.\ref{corr}, suggest a cubic $z^3_{1}$-behavior:
\begin{equation}
\varphi_S(z_1) \approx c(T) z_1^3 ~, \label{singul}
\end{equation}
for every  $T<T_{BKT}$, and for the values of $m$ smaller than the mean. Moreover, $c(T_{BKT})=0$.

\paragraph*{The large-$m$ tail at the BKT point. --}
Instead of using multicanonical Monte-Carlo simulations \cite{Berg} which 
need too large system sizes to conclude \cite{Hilfer},
we consider static in-plane magnetic field, $\bf H$, as a probe to study the 
features of the PDF for the large values of the magnetization. Indeed, as the intensity of $\bf H$ 
increases, the most probable magnetization, $m_H^{\star}$,
as well as its mean value, $\langle m_H \rangle$, explores larger
values of the PDF tail.
We consider two alternative forms for the critical tail, namely: 

$\bullet$ the `Gumbel-like' shape \eqref{bramwell} -- noted below: `~hypothesis $(G)$' --, 
which writes in the first-scaling form:
\begin{equation}
\Phi_{c}(z_{1}) \sim  
\exp \left(  - \lambda_{0} e^{a_{0} z_{1}} \right)  
\hskip 0.75cm \mbox{for} \hskip 0.75cm z_{1} \rightarrow \infty ~  ,~~~~(G) \notag
\end{equation}
with $a_0=a_{\sigma}\times(\langle m \rangle_c/\sigma_c)$ ($ \approx 16.4$ from \eqref{asigma}
and Table ~\ref{table1}), and 
$\lambda_0=\lambda_{\sigma}e^{-a_0}$. It is the form suggested 
in \cite{Bram1, Bram2, Bram3}; 

$\bullet$ the `Weibull-like' critical shape -- noted below: 
`~hypothesis $(W)$' --, which is \cite{book}:
\begin{equation}
\Phi_c(z_1) \sim 
\exp(-\lambda_1 z_1^{\delta+1}) \hskip 0.75cm \mbox{for} \hskip 0.75cm z_1 \rightarrow \infty ~  ,~~~~(W) \notag
\end{equation}
with $\lambda_1$ a positive parameter, and $\delta +1=16$ \cite{KosterlitzH}.\\

Let $\phi$ be the
direction of $\bf H$ with respect to the $x$-axis ({\it i.e.} ${\bf H}=(H\cos \phi,H\sin \phi)$).
According to general thermodynamics, the magnetization PDF  
is given by: $P_c(m,H) \propto \exp(-\beta_c {\cal F} + \beta_c L^2 m H \cos(\psi-\phi))$, 
with the field-less free energy $\cal F$. Therefore, the most probable 
magnetization, $m_H^{\star}$, is the solution of the equation $\partial P_c(m,H)/\partial m = 0$
for a given value of $H$. 
As the instantaneous
magnetization direction $\psi$ should coincide with the magnetic field direction $\phi$ for the large systems,
we use $\cos (\psi-\phi) \approx 1$.  
Rewritten in terms of  the auxiliary variables
$X \equiv H/ \langle m \rangle_c^{\delta}$ and $Y \equiv H/m_H^{\star \,\delta}$, 
Eqs.\eqref{free},\eqref{correction}, with 
hypothesis $(G)$ or $(W)$, result respectively in:
\begin{eqnarray}
\frac{X}{A}+ 2 b_1 \left( \frac{X}{Y} \right)^{1/\delta}  & = & 
\lambda_0 a_0 \, e^{a_0 (X/Y)^{1/\delta }}  \label{HG} \\
& \mbox{or} & \notag \\
& = & \lambda_1 (\delta + 1) \, \frac{X}{Y} ~, \label{HGW}
\end{eqnarray}
which are implicit equations for the 
most probable magnetization, $m_H^{\star}$,
(written in the combination $Y$) vs the magnetic field $H$
and the system size $N$ (written in the combination $X$).
The constant $A$ is such that: 
$A^{-1}=\beta_c L^2 \langle m \rangle_c^{\delta+1} \approx 1.07$.

For the large magnetic field, $m^\star_H$ is expected to be
much larger than  $\langle m \rangle_c$, that is: $X/Y \gg 1$. Consequently, the solution 
of Eq.\eqref{HG} is:
\begin{equation}
Y = a_0^{\delta}~ X / (\ln X+C)^{\delta} \label{soluceG}
\end{equation}
where $C = -\ln(A\lambda_0 a_0) \approx 14.2$ is a positive constant. 
\begin{figure}[h]
\includegraphics[angle=-90,width=1.2\columnwidth]{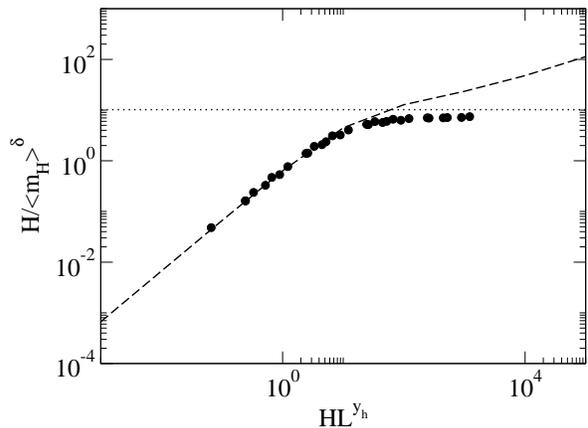}
\caption{\label{RH} Double-logarithmic plot of $H/\langle m_H \rangle^{\delta}$
{\it vs} the reduced magnetic field $HL^{y_H}$ with: $y_H=2\delta /(\delta +1)$.
These two variables are convenient for the numerical simulations and simply related to the variables $X$ and $Y$
of the text:
 $H/\langle m_H \rangle^{\delta} = Y \times (m_H^{\star}/\langle m_H \rangle)^{\delta}$
 and $HL^{y_H} = X \times (A \beta_c)^{-\delta/(\delta+1)}$, with the constant values:
$(m_H^{\star}/\langle m_H \rangle)^{\delta} \approx 1$ and $(A \beta_c)^{-\delta/(\delta+1)} \approx 0.96$.
The dashed curve is the solution of the first Eq.\eqref{HG} (corresponding to 
the hypothesis $(G)$), while the dotted line
is Eq.\eqref{soluceW}, with $\lambda_1=\lambda_{\sigma}$
in agreement with \eqref{l1ls}. The system
size goes from $L=16$ up to $L=512$. Each point corresponds
to an average over $100,000$ independent realizations \cite{footnote}.}
\end{figure}
Within the hypothesis $(W)$, one has: $ \left( X/Y \right)^{1/\delta} \ll  X/Y$,
such that \eqref{HGW} shows that $Y$ is asymptotically a constant:
\begin{equation}
Y \approx A \lambda_1(\delta+1) ~ . \label{soluceW}
\end{equation}
So, increase of $Y$ with the intense magnetic field should be 
interpretated as failure of $(W)$.

\subparagraph*{Inference from the numerical simulations. --}
Both solutions, \eqref{soluceG} and \eqref{soluceW}, are drawn on FIG.\ref{RH} in comparison
with the results of large-scale numerical simulations of the 2D XY-model with the in-plane magnetic field
at the BKT temperature. 
It is clear that 
the numerical simulations are consistent with the hypothesis $(W)$, while the
double-exponential tail $(G)$ should be discarded. This suggests the following form of the critical
PDF for the 2D XY-model:
\begin{equation}
P_c(m) \propto e^{\,b_1z_1^2-{\lambda }_{1} z_1^{16}} ~~, ~~ z_1 \equiv m/\langle m \rangle ~. \label{final}
\end{equation}
Below the BKT critical temperature, additional term $+c(T)z_1^3$ should appear  in the exponential. \\

In order to understand the origin of the approximation \eqref{bramwell}, let us change the
reduced magnetization according to: $z_1= 1+z_{\sigma}/(\langle m \rangle/\sigma)$. At $T_{BKT}$, and for the 
small values of 
$z_{\sigma}/(\langle m \rangle_c/\sigma_c)$ (recall that  
$\langle z_{\sigma} \rangle = 0$, and that $\langle m \rangle_c/\sigma_c \approx 14.8$ is a rather
large number), we obtain:
\begin{equation}
P_c(m) \propto e^{\,2b_1 z_{\sigma}/(\langle m \rangle_c/\sigma_c)-
{\lambda }_{1} (1+z_{\sigma}/(\langle m \rangle_c/\sigma_c))^{16}} ~. \notag
\end{equation}
Writing then 
$1+z_{\sigma}/(\langle m \rangle_c/\sigma_c) \approx e^{z_{\sigma}/(\langle m \rangle_c/\sigma_c)}$, yields
Eq.\eqref{bramwell}, provided the following relations are verified:
\begin{equation}
a_{\sigma} = \frac{16}{\langle m \rangle_c/\sigma_c} ~~; ~~
b_{\sigma} = \frac{2b_1}{\langle m \rangle_c/\sigma_c} ~~;~~
\lambda_{\sigma} = \lambda_1 ~, \label{l1ls}
\end{equation}
So, Eq.\eqref{bramwell} appears to be a good approximation around the most probable magnetization,
but is inconsistent with the general critical relation $\langle m_H \rangle \propto H^{1/\delta}$,
unlike Eq.\eqref{final}.
By the way, the conjectured relation \cite{Bram2} $b_{\sigma}/a_{\sigma} = \pi/2$ writes simply:
$b_1=4\pi$, that we accept here as a new conjecture (numerically: $b_1 \approx 12.7$, 
see FIG.\ref{TKT}).

\begin{table}
\caption{Temperature, system size, average magnetization per spin,
ratio of average magnetization to standard deviation. The best fit for 
the latter is: $\langle m \rangle_c/\sigma_c = 14.81-21.5/L$, at the BKT temperature, $T_{BKT}=0.893$.
\label{table1}}
\centerline{
\begin{tabular}{|c|c|c|r|} \hline\hline
$T$ & $L$  & $\langle m \rangle$ & $\langle m \rangle /\sigma$ \\ \hline
0.3 & 16 & 0.923218 & 66.958  \\
0.6 & 16 & 0.836307 & 29.249  \\
0.8 & 16 & 0.764091 & 18.260  \\
0.885 & 16 & 0.723259 & 13.907 \\ \hline
0.893 & 16 & 0.718814 & 13.467 \\  
0.893 & 32 & 0.662819 & 14.119 \\  
0.893 & 64 & 0.611181 & 14.486 \\  
0.893 & 96 & 0.582217 & 14.583 \\  
0.893 & 128 & 0.563209 & 14.644 \\ 
0.893 & 256 & 0.518921 & 14.687 \\ 
0.893 & 512 & 0.478045 & 14.829 \\ \hline\hline 
\end{tabular}}
\end{table}

\subparagraph*{Conclusion. --}
In this Letter, we explained how using the field conjugated to the
order parameter provides unique information about the tail of the probability
distribution function of the order parameter. This is of major importance 
for the critical systems, since the shape of the tail 
is directly linked to the value of a critical exponent. Therefore, this general method
provides alternative way to calculate or measure the critical exponent $\delta$.

We chose the critical 2D XY-model as a debated example to treat with this method.
Indeed, a former double-exponential approximation of the magnetization PDF in the 0-magnetic field 
is found to be inconsistent with the
critical behavior of the system - though 
correct near the most probable magnetization -. Moreover, this approximation being taken from
another field of statistical physics, could mislead, as it suggests hidden
link between these two fields. The new proposed approximation corrects 
these flaws.

\subparagraph{Acknowlegments. --}
The authors thank CNRS and FONACIT (PI2004000007) for their support.


\begin{thebibliography}{99}

\bibitem{Kolmogorov} a recent instance is: R. Botet and M. P{\l}oszajczak, Phys. Rev. Lett. {\bf 95}, 185702 (2005).
\bibitem{Bouchaud} J.-P. Bouchaud and A. Georges, Physics Reports {\bf 195}, 128 (1990).
\bibitem{Yang-Lee} C.N. Yang and T.D. Lee, Phys. Rev. {\bf 87}, 404 (1952);
T.D. Lee and C.N. Yang, Phys. Rev. {\bf 87}, 410 (1952).
\bibitem{Bram1} S.T. Bramwell {\it et al}, Phys. Rev. Lett. {\bf 84}, 3744 (2000).
\bibitem{Bram2} S.T. Bramwell {\it et al}, Phys. Rev. E {\bf 63}, 041106 (2001).
\bibitem{Bram3} G. Palma, T. Meyer and R. Labb\'e, Phys. Rev. E {\bf 66}, 026108 (2002).
\bibitem{holds} B. Portelli and P.C.W. Holdsworth, J. Phys. A {\bf 35}, 1231 (2002). 
\bibitem{extremes} R.D. Reiss and M. Thomas,  {\it Statistical Analysis of Extremal
Values}, Birkh\"auser (1997).
\bibitem{Bere} V.I. Berezinskii, Soviet Phys JETP {\bf 34}, 610 (1971).
\bibitem{KT}J.M. Kosterlitz and D.J. Thouless, J. Phys. {\bf C 6}, 1181 (1973).
\bibitem{nature} S.T. Bramwell, P.C.W. Holdsworth and J.-F. Pinton, Nature (London) {\bf 396}, 552 (1998) .
\bibitem{nous} R. Botet, M. P{\l}oszajczak and V. Latora, Phys. Rev. Lett. {\bf 78}, 4593 (1997).
\bibitem{FFS} J. Cardy (ed.), {\it Finite-Size Scaling}, North-Holland, Amsterdam (1988). 
\bibitem{Wolff} U. Wolff, Phys. Rev. Lett. {\bf 62}, 361 (1989).
\bibitem{Widom} B. Widom, J. Chem. Phys. {\bf 43}, 3898 (1965).
\bibitem{Amit} D.J. Amit, Y.Y. Goldschmidt and G. Grinstein, J. Phys. {\bf A 13}, 585 (1980).
\bibitem{Gulacsi} Z. Gul\'acsi and M. Gul\'acsi, Adv. in Physics {\bf 47}, 1 (1998).
\bibitem{Berg} B.A. Berg and T.Neuhaus, Phys. Rev. Lett. {\bf 68}, 9 (1992).
\bibitem{Hilfer} R. Hilfer {\it et al}, Phys. Rev. E {\bf 68}, 046123 (2003).
\bibitem{Berche} B. Berche, A. I. Fari{\~n}as and R. Paredes, Europhys. Lett. {\bf 60}, 539 (2002).
\bibitem{book} R. Botet and M. P{\l}oszajczak, {\it Universal Fluctuations}, World Scientific Lecture Notes in Physics,
New Jersey (2002).
\bibitem{KosterlitzH} J.M. Kosterlitz, J. Phys. {\bf C 7}, 1046 (1974).
\bibitem{footnote} For the 2D XY-model in the magnetic field $\bf H$, we  used a Wolff algorithm similar
to the one used for the ${\bf H}=0$ case, with an additional spin which interacts with all the other spins,
with $\bf H$ as the strength of the interaction. 
  
\end{thebibliography}
\end{document}